\documentclass[a4paper, 10 pt, conference]{ieeeconf}  

\IEEEoverridecommandlockouts                              

\usepackage{amssymb,amsmath,graphicx,multicol,epstopdf}

\newtheorem{Remark}{Remark}[section]

\newcommand{\N}{{\cal N}}
\newcommand{\R}{{\mathbb R}}

\title{\LARGE \bf
Regularized system identification using orthonormal basis functions
}

\author{Tianshi Chen and Lennart Ljung
\thanks{This research has been partially supported by
the Linnaeus Center CADICS, funded by the Swedish Research Council,
and the ERC advanced grant LEARN, No. 267381, funded by the European
Research Council, as well as by a research grant for junior
researchers, No. 2014-5894, funded by Swedish Research Council.}
\thanks{Tianshi Chen and Lennart Ljung are with the Department of Electrical Engineering,
Link\"{o}ping University, Link\"{o}ping, SE-58183, Sweden
        {\tt\small \{tschen,ljung\}@isy.liu.se}}%
}

\begin{document}

\maketitle
\thispagestyle{empty}
\pagestyle{empty}

\begin{abstract}
Most of existing results on regularized system identification focus
on regularized impulse response estimation. Since the impulse
response model is a special case of orthonormal basis functions, it
is interesting to consider if it is possible to tackle the
regularized system identification using more compact orthonormal
basis functions. In this paper, we explore two possibilities. First,
we construct reproducing kernel Hilbert space of impulse responses
by orthonormal basis functions and then use the induced reproducing
kernel for the regularized impulse response estimation. Second, we
extend the regularization method from impulse response estimation to
the more general orthonormal basis functions estimation. For both
cases, the poles of the basis functions are treated as
hyper-parameters and estimated by empirical Bayes method. Then we
further show that the former is a special case of the latter, and
more specifically, the former is equivalent to ridge regression of
the coefficients of the orthonormal basis functions.


\end{abstract}

\section{INTRODUCTION}

In this paper, we consider the system identification problem of
linear discrete-time, time-invariant and causal systems, which is
described as follows: \begin{align}\label{eq:dyn} y(t) = g^0*u(t) +
v(t), \quad t=1,\cdots,N
\end{align} where $t=1,\cdots,N$ are time indices at
which the measured input $u(t)$ and output $y(t)$ are collected, and
uniform sampling is used and the sampling interval $T_s=1$, $v(t)$
is the disturbance and for convenience assumed to be a zero mean
white Gaussian noise, $g^0(t)$ with $t=1,2,\cdots,$ is the impulse
response, $g^0*u(t)$ is the convolution of $g^0(t)$ and $u(t)$
evaluated at the time $t$. The goal is to estimate $g^0(t)$ as well
as possible based on the collected data $\{y(t),u(t)\}_{t=1}^N$.

The traditional method to this problem is the maximum
likelihood/prediction error method (ML/PEM), see e.g.,
\cite{Ljung:99,SoderstromS:89}. Since $v(t)$ is white, PEM first
postulates the so-called output error (OE) model structure
$G(q,\theta)$ with $\theta\in\R^n$:
\begin{align}\label{eq:model}
y(t) = G(q,\theta)u(t) + v(t),
\end{align}
where $q$ is the forward-shift operator and $qu(t)=u(t+1)$ and
\begin{align}\label{eq:oe} G(q,\theta)=\frac{B(q)}{F(q)},\begin{array}{c}\begin{aligned}
                                              B(q) &= b_1 q^{-1} +
\cdots + b_{n_b} q^{-n_b} \\
                                              F(q) &= 1+ f_1 q^{-1} +
\cdots + f_{n_f} q^{-n_f}
                                            \end{aligned}\end{array}
\end{align} with $  \theta = [b_1,\cdots, b_{n_b} f_1,
\cdots,f_{n_f}]^T$ and $ n=n_b+n_f $. As long as a model structure
$G(q,\theta)$ is chosen, ML/PEM minimizes the prediction error to
get the model estimate
\begin{align}\label{eq:pem/oe}
\hat\theta = \arg\min_{\theta} \sum_{t=1}^N (y(t) -
G(q,\theta)u(t))^2.
\end{align} Since the disturbance $v(t)$ in (\ref{eq:dyn}) is
modeled as a stochastic process, the estimate $\hat\theta$ is a
random variable. Let $\hat g$ denote the impulse response of
$G(q,\hat\theta)$. Then the mean square error (MSE) $\mathbb
E(\|\hat g-g^0\|_2^2)$ tells the quality of the estimate
$\hat\theta$. For the chosen model structure $G(q,\theta)$, a key
issue to reduce the MSE is to find the ``right'' model complexity:
it shall be parsimonious but capable to describe the data.
Traditionally, it is suggested to use the model structure selection
criterion, like AIC, BIC, to find a suitable $n$, the dimension of
$\theta$. However, this way may not work well for short and noisy
data records.

The model structure (\ref{eq:oe}) has a very general form and
includes many widely used model structures as special cases. One
attractive class of model structures among many others is the
linear-in-parameter model structures which can considerably simplify
the optimization in (\ref{eq:pem/oe}). The most well-known instance
is perhaps the finite impulse response (FIR) model structure
\begin{align}\label{eq:fir}   G(q,\theta) = \sum_{k=1}^{n} g_k
q^{-k},\quad \theta = [g_1,\cdots, g_n]^T.
\end{align}
However, the FIR model is often criticized for its large variance
error when high order FIR models have to be used to describe
``slow'' systems with either slow dynamics or with high sampling
rate. A more compact model structure is the linear combination of
basis functions:
\begin{align} \label{eq:ob} G(q,\theta) = \sum_{k=1}^{m} g_k \bar
F_k(q), \theta = [g_1,\cdots, g_m, f_1,\cdots, f_{n_f}]^T
\end{align} where $n=m+n_f$ and $\bar F_k(q)=q^{k-1}/F(q)$, $k=1,\cdots,m$ are pre-specified basis functions. The
model structure (\ref{eq:ob}) has attracted a lot of interests in
the last two decades, see e.g., \cite{HVW05book} and the references
therein. Two widely known special cases of (\ref{eq:ob}) are the
Laguerre model \cite{Wahlberg:91} and the Kautz model
\cite{Wahlberg94}. The Laguerre model takes the form
\begin{align}\label{eq:lag}
G(q,\theta) &= \sum_{k=1}^{m} g_k
\frac{\sqrt{(1-a^2)}}{q-a}\left(\frac{1-aq}{q-a}\right)^{k-1}\\\nonumber
&\theta = [g_1,\cdots, g_m, a]^T, \quad |a|<1
\end{align} where $a$ is pole of the Laguerre model and has to be pre-specified according
to the \emph{a priori} information on the time constant of the
underlying system \cite{Wahlberg94}. Since the basis functions have
infinite impulse responses, there is often no problem of describing
``slow'' systems with relatively small number of basis functions in
(\ref{eq:ob}). While the use of orthonormal basis functions
(\ref{eq:ob}) has been discussed a lot, still open problems are
\begin{enumerate}
\item how to choose suitable poles for the basis function?
\item how many basis functions shall be used?
\end{enumerate}

There is another way to reduce the MSE, i.e, by using
regularization. However, this way has not been investigated
rigorously in system identification until the seminal work
\cite{PN10a}. Instead of trimming the model complexity of
$G(q,\theta)$ in terms of $n$, it was suggested to use a well-tuned
regularization to regularize the impulse response to reduce the MSE
\cite{COL12a}. Since then the followup results in
\cite{PCN11,COL12a,CALCP14,PC14,CCLP14} and the recent survey paper
\cite{PDCDL14} show that the regularized high order FIR model (or
high order ARX model) can lead to good model estimates in terms of
accuracy and robustness. In this paper, we will make use of
orthonormal basis functions for the regularized system
identification and we will consider two cases. First, we construct
reproducing kernel Hilbert space of impulse responses by orthonormal
basis functions and then use the induced reproducing kernel for the
regularized impulse response estimation. Second, we extend the
regularization method from impulse response estimation to the more
general orthonormal basis functions estimation. For both cases, the
poles of the basis functions are treated as hyper-parameters and
estimated by empirical Bayes method. Then we further show that the
former is a special case of the latter, and more specifically, the
former is equivalent to ridge regression of the coefficients of the
orthonormal basis functions.

\section{Regularized least squares method}\label{sec:rls}

Consider a linear regression model
\begin{align}
\label{eq:lrg}
  Y_N=\Phi_N\theta+V_N
\end{align}
where $Y_N\in\R^N$ is the data, $\Phi_N\in\R^{N\times n}$ is the
regression matrix, $\theta\in\R^n$ is the parameter to be estimated,
and $V_N$ is the disturbance and assumed to be white Gaussian
distributed as $\N(0,\sigma^2I_N)$ with $I_N$ being the
$N$-dimensional identity matrix. We estimate $\theta$ by minimizing
the regularized least squares (RLS) criterion
\begin{subequations}\label{eq:thetae}
\begin{align}\label{eq:rls} \hat\theta & 
= \arg\min_\theta \|Y_N-\Phi_N\theta\|_2^2 +
\sigma^2\theta^T\textbf{K}(\alpha)^{-1}\theta\\&=K
(\alpha)\Phi_N^T(\Phi_N\textbf{K} (\alpha)\Phi_N^T+\sigma^2I_N)^{-1}
Y_N.\label{eq:theta_expression}
\end{align}
\end{subequations}
Here, $\textbf{K}(\alpha)\succeq0$\footnote{When
$\textbf{K}(\alpha)$ is singular, (\ref{eq:rls}) has to be
interpreted in the way discussed in \cite[Remark 2.1]{CALCP14}. } is
called the \emph{regularization} matrix (also often called the
kernel matrix) and defined through the kernel function
$K(k,j;\alpha)$ as $\textbf{K}_{k,j}(\alpha)=K(k,j;\alpha)$, where
$\alpha$ is a vector of tuning parameters and called
hyper-parameter.

There are two key issues: \begin{enumerate}
\item how to parameterize the kernel function $K(k,j;\alpha)$ which is often simply written as
$K(\alpha)$ below?
\item how to tune the hyper-parameter $\alpha$?
\end{enumerate}
For 1), it is worth to note \cite[Theorem 1]{COL12a} that the
optimal regularization matrix in the sense of minimizing the MSE
matrix of $\theta$ with respect to $\theta_0$ (the true value of
$\theta$) exists and takes the form of
$\textbf{K}^{Opt}=\theta_0\theta_0^T $. While it cannot be applied
in practice, it gives a guideline to design the regularization
matrix: let it mimic the behavior of $\textbf{K}^{Opt}$. Apparently,
if some prior information is known for $\theta_0$, it shall be used
in the design of a suitable kernel function $K(\alpha)$.

For 2), the current most effective method is to embed the
regularization in the Bayesian framework and invoke the empirical
Bayes method, i.e., the marginal likelihood maximization. Assume
$\theta\sim\N(0,\textbf{K}(\alpha))$. Then we estimate $\alpha$ by
maximizing
\begin{align}\nonumber
\hat\alpha &= \arg\max_\alpha p(Y_N|\alpha)\\
 &= \arg\min_\alpha Y_N^T(\Phi_N\textbf{K}(\alpha)\Phi_N^T\nonumber\\ &+
 \sigma^2I_N)^{-1}Y_N + \log\det(\Phi_N\textbf{K}(\alpha)\Phi_N^T + \sigma^2I_N)
\label{eq:margLmax}\end{align}

\subsection{Regularized impulse response estimation}

For regularized impulse response estimation, we consider the model
(\ref{eq:fir}) with $n=\infty$. The system (\ref{eq:model}) can then
be written as a linear regression model (\ref{eq:lrg}) with the
$i$th row of $Y_N,V_N$ and $\Phi_N$ being $y(i),v(i)$ and
$\varphi(i)=[u((i-1)),\cdots,u((i-\infty))]^T$ where the unknown
inputs $u(t)$ are set to zero, and $\theta=[g_1,g_2,\cdots,
]^T\in\R^{\infty}$. So we can use the RLS method to estimate the
impulse response. The remaining issue is the design of a suitable
kernel function $K(\alpha)$. Several choices have been suggested in
\cite{PN10a,PCN11,COL12a}. For example, the diagonal-correlated (DC)
kernel and its special case, the tuned-correlated (TC) kernel are
defined as:
\begin{align}\label{eq:DC} &\text{DC}
&K^{dc}(k,j;\alpha)&=c\lambda^{(k+j)/2}\rho^{|k-j|}, \alpha = [c \
\lambda\ \rho]^T \\ \label{eq:TC} &\text{TC}
&K^{tc}(k,j;\alpha)&=c\min(\lambda^k,\lambda^j), \alpha = [c\
\lambda]^T
\end{align} where the TC kernel has also been introduced as the
first-order stable spline (SS) kernel, see \cite{PN11l,COGL11l} for
discussions. In practice, we however cannot handle infinite impulse
response and we have to truncate the infinite impulse response to a
finite one, i.e., the FIR model. In this case, we refer this method
as the regularized FIR model method in \cite{COL12a}.

\section{Regularized impulse response estimation with kernel structure constructed by orthonormal basis functions
}\label{sec:obk}

In the following, we consider a different kernel which is
constructed by use of the orthonormal basis functions. Before
proceeding to the details, recall that the RLS criterion
(\ref{eq:rls}) for regularized impulse response estimation has a
function estimation interpretation. The RLS (\ref{eq:rls}) is
equivalent to
\begin{align}
\label{eq:feinrkhs} \hat\vartheta & 
= \arg\min_{\vartheta\in{\mathcal H_{K(\alpha)}}}
\sum_{t=1}^N|y(t)-\vartheta*u(t)|^2 +
\sigma^2\|\vartheta\|^2_{\mathcal H_{K(\alpha)}}
\end{align} where $\vartheta(t)=\theta_t$ with $\theta_t$ being
the $t$th element of $\theta\in\R^{\infty}$ is the impulse response,
and $\mathcal H_{K(\alpha)}$ is the reproducing kernel Hilbert space
(RKHS) induced by the kernel $K(\alpha)$. Then the RLS estimate is
also the function estimate that minimizes (\ref{eq:feinrkhs}) within
the RKHS $\mathcal H_{K(\alpha)}$. This implies that when we trim
the kernel $K(\alpha)$, we equivalently trim the function space
where we search for the impulse response.

The above observation gives us another idea to design the kernel
structure: we can first construct a RKHS space of suitable impulse
responses and this space then uniquely determines a reproducing
kernel according to \emph{Moore-Aronszajn} Theorem, see e.g.,
\cite{Aronszajn50}. Note that looking for a RKHS space of impulse
responses in time domain is equivalent to looking for a RKHS space
of transfer functions in frequency domain. In system identification
community, the idea of approximating or expressing the transfer
function of the underlying system by expanding it in terms of
orthogonal basis functions have been well studied, see e.g.,
\cite{HVW05book,HVB95,VHB95,NHG99,NinnessHG:99a} and the references
therein. It is natural to ask if the space spanned by orthogonal
basis functions could be a candidate for our use. To answer this
question, we have to check if this space is a RKHS, and if it is,
what its reproducing kernel is. Fortunately, there are standard
answers to these questions.

\subsection{Transfer function space spanned by the orthonormal basis functions on the unit circle \cite{DjrBashian66,NHG99}}


Following \cite{DjrBashian66}, let $\{\alpha_k\}_{k=0}^\infty$ with
$|\alpha_k|<1$ be an arbitrary sequence of complex numbers which may
appear as numbers of finite or even infinite multiplicity. Given
$\{\alpha_k\}_{k=0}^\infty$, a system of functions
$\{\phi_k(e^{i\omega})\}_{k=0}^\infty$ is defined as
\begin{equation}\label{eq:orthobasis}
\begin{aligned}
\phi_0(e^{i\omega}) &= \frac{\sqrt{1-|\alpha_0|^2}}{1-\overline{\alpha_0}e^{i\omega}},\\
\phi_j(e^{i\omega}) & =
\frac{\sqrt{1-|\alpha_j|^2}}{1-\overline{\alpha_j}e^{i\omega}}\prod_{k=0}^{j-1}\frac{\alpha_k-e^{i\omega}}{1-\overline{\alpha_k}
e^{i\omega}}\frac{|\alpha_k|}{\alpha_k}, \ j=1,2,\cdots,
\end{aligned}
\end{equation}
where $\overline{\alpha_j}$ means the complex conjugate of
$\alpha_j$, $\omega\in[-\pi\ \pi)$, and
$\frac{|\alpha_j|}{\alpha_j}=\frac{\overline{\alpha_j}}{|\alpha_j|}=-1$
for $\alpha_j=0$. Such a system is called the \emph{Malmquist
system}. It is well-known that the Malmquist system is orthonormal
on the unit circle in the sense that
\begin{align}\label{eq:oborthonormal} \frac{1}{2\pi} \int_{-\pi}^\pi
\phi_k(e^{i\omega})\overline{\phi_j(e^{i\omega})} d\omega
=\delta_{k,j}= \left\{\begin{array}{cc}
                                                         0 & k\neq j \\
                                                         1 & k=j
                                                       \end{array}
\right.
\end{align}

We are interested in the space spanned by a subset of the Malmquist
system (\ref{eq:orthobasis}). It can be shown see e.g., \cite{NHG99}
that the space spanned by
$\{\phi_0(e^{i\omega}),\phi_1(e^{i\omega}),\cdots,\phi_m(e^{i\omega})\}$
with the inner product defined as \begin{align}  \langle f, g
\rangle = \frac{1}{2\pi}\int_{-\pi}^\pi
f(e^{i\omega})\overline{g(e^{i\omega})} d\omega
\end{align} is a RKHS space with the reproducing kernel
\begin{align}\label{eq:obkernel} K_{freq}^{ob}(e^{i\omega},e^{i\omega'}) = \sum_{k=0}^m
\phi_k(e^{i\omega})\overline{\phi_k(e^{i\omega'})}
\end{align} which we will refer below as the $(m+1)$th order orthonormal
basis (OB) kernel in frequency domain.

Setting in the following \begin{align}   B_{m+1}(e^{i\omega}) =
\prod_{k=0}^{m}
\frac{\alpha_k-e^{i\omega}}{1-\overline{\alpha_k}e^{i\omega}}
\frac{|\alpha_k|}{\alpha_k}
\end{align} where same as before, $\frac{|\alpha_k|}{\alpha_k}=\frac{\overline{\alpha_k}}{|\alpha_k|}=-1$
for $\alpha_k=0$. Then the kernel (\ref{eq:obkernel}) has a
simplified expression \cite[Lemma 5]{DjrBashian66}
\begin{align}\label{eq:obkernel_cd}
K_{freq}^{ob}(e^{i\omega},e^{i\omega'})
=\frac{1-B_{m+1}(e^{i\omega})\overline{B_{m+1}(e^{i\omega'})}}{1-e^{i(\omega-\omega')}}\end{align}
which is also known as the Christoffel-Darboux (C-D) formula, see
e.g., \cite[Theorem 3.1]{NHG99}. The C-D formula is useful to
simplify the construction of the kernel matrix (i.e., the
regularization matrix).

\subsection{The Laguerre kernel}

The simplest case of OB kernels (\ref{eq:obkernel_cd}) is perhaps
the case where $\alpha_i=a$ for $i=0,1,\cdots,m$ with $a\in\R$ and
$|a|< 1$. In this case, the OB kernel (\ref{eq:obkernel_cd}) becomes
\begin{align}\label{eq:lagkernel}K_{freq}^{lag}(e^{i\omega},e^{i\omega'}) = \left\{\begin{array}{cc}
                        \frac{1-B_{m+1}(e^{i\omega})\overline{B_{m+1}(e^{i\omega'})}}{1-e^{i(\omega-\omega')}} & \omega\neq \omega' \\
                        (m+1)\frac{1-a^2}{|1-ae^{i\omega}|^2} & \omega =
                        \omega'
                      \end{array}\right.
\end{align}
which we will refer below as the $(m+1)$th order  Laguerre kernel.
This is because it is the reproducing kernel of the RKHS space
spanned by the first $m+1$ Laguerre rational basis function of
(\ref{eq:lag}) in the frequency domain. For the Laguerre kernel
(\ref{eq:lagkernel}), there is only one hyper-parameter $a$, the
real pole of the Laguerre basis functions, which is convenient to
estimate for the hyper-parameter estimation.

\subsection{Regularized frequency response estimation}

Since OB kernels are defined in frequency domain, one may wonder if
it is possible to work in frequency domain directly without going
back to the time domain. The answer is affirmative. Recently, we
have derived the dual of the regularized impulse response estimation
in frequency domain, i.e., the regularized frequency response
estimation, see \cite{LC14} for details. By using the implementation
in \cite{LC14}, we can derive the regularized frequency response
with OB kernels.

\section{Regularized orthonormal basis functions estimation } \label{sec:rob}

It is worth to note that the hyper-parameters of OB kernels
(\ref{eq:obkernel_cd}) are $\{\alpha_k\}_{k=0}^m$ which are the
poles of the basis functions. So tuning OB kernels is equivalent to
tuning the location of the poles of its underlying basis functions.
This finding motivates another way of using orthonormal basis
functions for regularized system identification:

\begin{enumerate}
\item formulate the orthonormal basis functions based model as a
linear regression model;

\item treat the poles of the orthonormal basis functions as
hyper-parameters and design a suitable kernel for the coefficients
of the orthonormal basis functions;

\item estimate the hyper-parameter by empirical Bayes method and
then obtain the regularized orthonormal basis functions by using RLS
method.

\end{enumerate}

We first formulate (\ref{eq:model}) with the linear combination of
orthonormal basis functions (\ref{eq:ob}) as a linear regression
model: \begin{align} \label{eq:lrg-ob}
  y(t)=\sum_{k=1}^mg_k
\varphi_k*u(t)+v(t)
\end{align}
where $g = [g_1,\cdots, g_m]^T$, $\varphi_k(t)$ is the impulse
response of $\bar F_k(q)$ in (\ref{eq:ob}). Let $p$ be the vector
consisting of all poles of $\bar F_k(q)=0$, $k=1,\cdots,m$. Then the
impulse response $\varphi_k(t)$, $k=1,\cdots,m$ depend on $p$.

The feature of OB kernels that their hyper-parameters are poles of
the basis functions motivates to treat poles of the basis functions
as hyper-parameters and estimate them by the empirical Bayes method.
It should be noted that this idea has also been figured out
independently by Darwish, T\'{o}th, and Van den Hof in \cite{DTV14}.
For now we assume that $p$ is known and then we can estimate $g$ by
minimizing the RLS criterion
\begin{align}
\label{eq:rls-ob}
 \hat g & 
= \arg\min_g \sum_{t=1}^N|y(t)-\sum_{k=1}^mg_k \varphi_k*u(t)|^2 +
\sigma^2g^T\textbf{K}(\alpha)^{-1}g\\
&= \arg\min_g \|Y_N-\Phi_N(p)g\|_2^2 +
\sigma^2g^T\textbf{K}(\alpha)^{-1}g \label{eq:rls-ob2}
\end{align}
where $\textbf{K}(\alpha)$ is the regularization matrix on the
coefficients $\{g_k\}_{k=1}^m$ of the orthonormal basis functions,
and $\Phi_N(p)$ is the regression matrix that can be formed in a
natural way.

As discussed in Section \ref{sec:rls}, it is a key issue to design a
suitable kernel structure, which relies on the prior knowledge that
we know about the coefficients of the orthonormal basis functions.
Apparently, this issue depends on what orthonormal basis functions
we use. For illustration, we consider the Laguerre model
(\ref{eq:lag}) as an example below.

The assumptions on the Laguerre coefficients $\{g_k\}_{k=1}^\infty$
and the convergence property of Laguerre model, i.e, how fast
(\ref{eq:lag}) converges as $m\rightarrow\infty$ has been discussed,
see e.g., \cite{Wahlberg:91}. It is suggested in \cite{Wahlberg:91}
to assume the absolute convergence of the sum of the Laguerre
coefficients $\{g_k\}_{k=1}^{\infty}$, i.e.,
\begin{align}\label{eq:assonLagcoe}
  \sum_{k=1}^\infty |g_k|<\infty
\end{align}
If we treat $\{g_k\}_{k=1}^\infty$ as the impulse response of a
linear system, then the above assumption (\ref{eq:assonLagcoe}) says
nothing but the linear system is stable. This observation implies
that the kernels introduced for regularized impulse response
estimation, the SS, TC and DC kernels can be candidates  to
regularize the Laguerre coefficients $\{g_k\}_{k=1}^\infty$.

\begin{Remark}
As pointed out in \cite{Wahlberg:91}, the convergence rate of the
Laguerre model (\ref{eq:lag}) can be slow, e.g., if the system has
poles close to the unit circle or has high resonant poles. In this
case, one can try the adapted DC kernel as follows:
\begin{align}
K^{adc}(k,j)&=c\lambda(k+j)\rho^{|k-j|},
\end{align} where $\lambda(\cdot)$ is a nonnegative function such that $\lambda(\cdot)$ decays
slower than the exponential function and $K^{adc}$ is a valid
kernel. Or one can choose to use the other regularized orthonormal
basis functions, such as the Kautz model in \cite{Wahlberg94} to
handle the case where the system has high resonant poles.
\end{Remark}

For more general orthonormal basis functions, we can always first
try the SS, TC and DC kernels if (\ref{eq:assonLagcoe}) is assumed.
If they do not work so well, we shall spend more efforts on
investigating the prior knowledge or assumption on the coefficients
of orthonormal basis functions and design a suitable kernel
structure accordingly.

Now it remains to estimate the hyper-parameters: the pole $p$ of the
orthonormal basis functions and the hyper-parameter $\alpha$ used to
parameterize the kernel structure. Assume
$\theta\sim\N(0,\textbf{K}(\alpha))$. Then from (\ref{eq:rls-ob2})
we have
\begin{align}\nonumber
\hat p, \hat\alpha &= \arg\max_{p,\alpha} p(Y_N|p,\alpha)\\
 &= \arg\min_{p,\alpha} Y_N^T(\Phi_N(p)\textbf{K}(\alpha)\Phi_N(p)^T\nonumber\\\nonumber &+
 \sigma^2I_N)^{-1}Y_N + \log\det(\Phi_N(p)\textbf{K}(\alpha)\Phi_N(p)^T + \sigma^2I_N)
\end{align}
Finally, solving (\ref{eq:rls-ob}) or (\ref{eq:rls-ob2}) by
replacing $p,\alpha$ with $\hat p,\hat\alpha$ yields the regularized
orthonormal basis function estimate.


\section{Regularized impulse response estimation with the OB kernel is a special case of regularized orthonormal basis functions
estimation}\label{sec:relation}

In this section, we show that the regularized impulse response
estimation with the OB kernel (\ref{eq:obkernel}) is a special case
of the regularized orthonormal basis functions estimation. More
specifically, it is equivalent to ridge regression of the
coefficients of the orthonormal basis functions, see e.g.,
\cite{Bishop06}.

To show this, it is more convenient to go back to time domain. For
the orthonormal basis functions
$\{\phi_k(e^{i\omega})\}_{k=0}^\infty$ in frequency domain, we can
define their correspondents $\{\varphi_k(t)\}_{k=0}^\infty$ in time
domain. Here, $\varphi_k(t)$ is the impulse response of
$\phi_k(e^{i\omega})$ and moreover, we have
\begin{align} \phi_k(e^{i\omega})=\mathcal F\{\varphi_k(t)\},\quad\varphi_k(t)=\mathcal F^{-1}\{\phi_k(e^{i\omega})\}
\end{align} where $\mathcal F$ and $\mathcal F^{-1}$ denote the
discrete time Fourier transform and its inverse transform,
respectively.

Then it is straightforward to verify by using
(\ref{eq:oborthonormal}) and Cauchy's integral formula that
$\{\varphi_k(t)\}_{k=0}^\infty$ are orthonormal in the sense that
\begin{align}
\sum_{t=0}^\infty \varphi_k(t)\varphi_j(t) =\delta_{k,j}=
\left\{\begin{array}{cc}
                                                         0 & k\neq j \\
                                                         1 & k=j
                                                       \end{array}
\right.
\end{align} Moreover,
the space spanned by
$\{\varphi_0(t),\varphi_1(t),\cdots,\varphi_m(t)\}$ with the inner
product\begin{align}  \langle f, h \rangle = \sum_{t=0}^\infty
f(t)h(t)
\end{align} is a RKHS space with the reproducing kernel
\begin{align}\label{eq:obkernel_time} K_{time}^{ob}(t,t') = \sum_{k=0}^m
\varphi_k(t)\varphi_k(t')
\end{align} which we will refer below as the $(m+1)$th order OB kernel in time domain.
Apparently, the OB kernel (\ref{eq:obkernel_time}) in time domain
and the OB kernel (\ref{eq:obkernel}) in frequency domain are
related through Fourier transform, e.g.,
$K_{freq}^{ob}(e^{i\omega},e^{i\omega'})=\mathcal F\{\mathcal
F\{K_{time}^{ob}(t,t')\}\}$.

Now consider (\ref{eq:feinrkhs}) with the kernel $K(\alpha)$
replaced by the OB kernel (\ref{eq:obkernel_time}). The RKHS
$\mathcal H_{K(\alpha)}$ becomes \begin{align}\nonumber \mathcal
H_{K(\alpha)} &= \text{ span of
$\varphi_0(t),\varphi_1(t),\cdots,\varphi_m(t)$}\\
&=\{\vartheta(t)|\vartheta(t)=\sum_{k=1}^m g_k
\varphi_k(t),g_k\in\R\}\end{align}  and moreover,
\begin{align} \|\vartheta\|^2_{\mathcal H_{K(\alpha)}} =
\sum_{k=1}^mg_k^2\end{align} Therefore, (\ref{eq:feinrkhs}) is
equivalent to \begin{align}
\label{eq:feinrkhs2} \hat g & 
= \arg\min_{g} \sum_{t=1}^N|y(t)-\sum_{k=1}^mg_k \varphi_k*u(t)|^2 +
\sigma^2\|g\|^2_2
\end{align} where the regularization $\|g\|^2_2$ is a ridge regression
of $g$.

We have the following interesting observations:
\begin{itemize}
\item[1)] The regularized impulse response estimation
with the OB kernel (\ref{eq:obkernel_time}) (equivalently,
(\ref{eq:obkernel})) is equivalent to a ridge regression of the
coefficients of the orthonormal basis functions
(\ref{eq:feinrkhs2}), which is a special case of the regularized
orthonormal basis functions estimation (\ref{eq:rls-ob}) with the
regularization matrix $\textbf{K}(\alpha)=I_m$.

\item[2)] For the Laguerre kernel (\ref{eq:lagkernel}), the ridge
regression $\|g\|^2_2$, i.e., the kernel
$K(k,j;\alpha)=\alpha\delta_{k,j}$ cannot guarantee the absolute
convergence of the sum of Laguerre model coefficients, i.e.,
(\ref{eq:assonLagcoe}). Since the kernel
$K(k,j;\alpha)=\alpha\delta_{k,j}$ does not reflect our prior
knowledge, it is not a good kernel and the regularized impulse
response estimation with the OB kernel (\ref{eq:obkernel_time}) will
not work well for high order OB kernel. This claim will be verified
by numerical simulations shortly.

\end{itemize}

\section{Numerical simulation}\label{sec:sim}

\subsection{Data-bank}

For this preliminary work, we use a portion of the data-bank in
\cite[Section 2]{COL12a}, which consists of 4 data collections:
\begin{itemize}
\item \texttt{S1D1}: fast systems, data sets with $N=500$, SNR=10
\item \texttt{S2D1}: slow systems, data sets with $N=500$, SNR=10
\item \texttt{S1D2}: fast systems, data sets with $N=375$, SNR=1
\item \texttt{S2D2}: slow systems, data sets with $N=375$, SNR=1
\end{itemize}
Each collection contains 250 randomly generated 30th order
discrete-time systems and data sets. The fast systems have all poles
inside the circle with center at the origin and radius 0.95 and the
slow systems have at least one pole outside this circle. The signal
to noise ratio (SNR) is defined as the ratio of the variance of the
noise-free output over the variance of the white Gaussian noise. In
all cases the input is Gaussian random signal with unit variance.
For more details regarding the data bank, see \cite[Section
2]{COL12a}.

\subsection{Examined methods}

We examine three methods:
\begin{enumerate}

\item \textbf{RLAG-TC,RLAG-DI}: the regularized Laguerre basis functions estimation. The
Laguerre model with orders $m=10,20,30,40$ are considered. The TC
kernel (\ref{eq:TC}) and the diagonal (DI) kernel
$K(k,j;\alpha)=\text{diag}(\alpha,\alpha^2,\cdots,\alpha^m)$ are
used to regularize the Laguerre coefficients. The results are
represented as RLAG-TC and RLAG-DI, respectively.

\item \textbf{LS-LAG}: the Laguerre basis function estimation with least squares
method. The estimate of the pole of the Laguerre model is obtained
from RLAG-TC and then the least squares method is used to estimate
the Laguerre coefficients without regularization.

\item \textbf{RFIR-TC,RFIR-LAG}: the regularized impulse response estimation. The order of the FIR
model (\ref{eq:fir}) is chosen to be 125 and the unknown input are
set to zero when forming the regression matrix. The TC kernel
(\ref{eq:TC}) and the Laguerre kernel (\ref{eq:lagkernel}) are used
to regularize the impulse response coefficients. The results are
represented as RFIR-TC and RFIR-LAG, respectively. As shown in
Section \ref{sec:relation}, RFIR-LAG is equivalent to regularized
Laguerre basis functions estimation with the scaled identity kernel
$K(k,j;\alpha)=\alpha\delta_{k,j}$.

\end{enumerate}

\subsection{Model fit}

To measure the performance of the examined methods, we compare the
impulse response of the estimated model with that of the true
system: we let $\hat g_k$ and $g_k^0$ to denote the $k$th
coefficient of the former and the latter impulse response,
respectively. Then the model fit is defined as
\small\begin{align}\label{eq:fit} W =100\left(1 -
\left[\frac{\sum^{125}_{k=1}|g_k^0-\hat{g}_k|^2
}{\sum^{125}_{k=1}|g_k^0-\bar{g}^0|^2}\right]^{1/2}\right),\quad
\bar{g}^0=\frac{1}{125}\sum^{125}_{k=1}g^0_k
\end{align}\normalsize

\subsection{Simulation result}

The average model fit over the corresponding data collections are
shown in the table below.

\begin{center}
\vspace{2mm}

\begin{tabular}{ccccc}
  \hline
  LS-LAG & \texttt{S1D1} & \texttt{S1D2} & \texttt{S2D1} & \texttt{S2D2} \\\hline
  $m=10$ & 80.2 & 70.2 & 73.6 & 59.5 \\

  $m=20$ & 88.8 & 68.6 & 82.1 & 56.3 \\
  $m=30$ & 90.0 & 62.8 & 84.0 & 38.1 \\
  $m=40$ & 88.7 & 56.6 & 84.1 & -4.9 \\
  \hline
\end{tabular}
\vspace{1mm}

\begin{tabular}{ccccc}
  \hline
  RFIR-TC & \texttt{S1D1} & \texttt{S1D2} & \texttt{S2D1} & \texttt{S2D2}
  \\\hline
$n=125$   & 91.4 & 76.1 & 81.2 & 66.1 \\
  \hline
\end{tabular}
\vspace{1mm}

\begin{tabular}{ccccc}
  \hline
  RFIR-LAG & \texttt{S1D1} & \texttt{S1D2} & \texttt{S2D1} & \texttt{S2D2} \\\hline
  $m=10$ & 80.1 & 69.6 & 72.0 & 60.6 \\
 $m=20$ & 88.3 & 68.5 & 80.4 & 61.3 \\
  $m=30$ & 89.1 & 64.7 & 82.5 & 59.9 \\
  $m=40$ & 88.1 & 62.6 & 83.1 & 58.4 \\
  \hline
\end{tabular}

\begin{tabular}{ccccc}
  \hline
  RLAG-TC & \texttt{S1D1} & \texttt{S1D2} & \texttt{S2D1} & \texttt{S2D2} \\\hline
  $m=10$ & 80.2 & 71.3 & 72.9 & 63.0 \\
 $m=20$ & 89.2 & 75.3 & 81.9 & 67.8 \\
  $m=30$ & 91.3 & 76.1 & 85.2 & 69.2 \\
  $m=40$ & \textbf{91.8} & \textbf{76.3} & \textbf{86.8} & \textbf{70.1} \\
  \hline
\end{tabular}
\vspace{1mm}

\begin{tabular}{ccccc}
  \hline
  RLAG-DI & \texttt{S1D1} & \texttt{S1D2} & \texttt{S2D1} & \texttt{S2D2} \\\hline
  $m=10$ & 80.4 & 71.8 & 73.2 & 64.0 \\
 $m=20$ & 89.2 & 75.7 & 82.3 & 68.6 \\
  $m=30$ & 91.2 & 76.0 & 85.9 & 69.7 \\
  $m=40$ & 91.6 & 76.1 & \textbf{86.8} & 70.0\\
  \hline
\end{tabular}
\vspace{1mm}

\end{center}


%
%

\subsection{Findings}

First, RLAG can achieve comparable performance as RFIR but with more
compact model stucture in terms of the number of basis functions. In
particular, for slow systems \texttt{S2D1} and \texttt{S2D2}, RLAG
has clearly better performance (about 5\%) than RFIR.

Second, for RLAG, RLAG-DI has very close performance as RLAG-TC,
which is different from RFIR studied in \cite{COL12a} where RFIR-DI
is much worse than RFIR-TC. For RFIR, TC kernel is clearly a better
kernel than the DI kernel because on the one hand, the impulse
response is often smooth and on the other hand, the latter does not
assume smoothness. However, for RLAG, no prior knowledge regarding
the Laguerre coefficients is available except the absolute
convergence of the sum of the Laguerre coefficients
(\ref{eq:assonLagcoe}). Both TC kernel and DI kernel can guarantee
(\ref{eq:assonLagcoe}). The simulation results indicate that to
assume independence between neighboring Laguerre coefficients is not
a bad choice for the tested data bank.

Third, RFIR-LAG has worse performance than RLAG. This coincides with
our observation in Section \ref{sec:relation} that the ridge
regression is not a suitable regularization for Laguerre basis
functions. The influence of the unsuitable regularization is
enlarged for high order Laguerre kernels and cause larger difference
in the performance.

Fourth, RLAG has better performance than LS-LAG shows the importance
of the regularization on the Laguerre coefficients.


\section{Conclusion and future works}

In this preliminary work, we have explored the possibilities to
tackle regularized system identification problems using orthonormal
basis functions.


Interestingly, the idea of constructing kernels using orthonormal
basis functions for regularized impulse response estimation turns
out to be a special case of the regularized orthonormal basis
functions estimation, and moreover, it is equivalent to ridge
regression of the coefficients of the orthonormal basis functions.

The idea of regularizing the orthonormal basis functions works fine
but still requires more careful investigation. Due to the space
limitation we have mainly studied the regularized Laguerre basis
functions as an instance, but the proposed idea applies to the more
general orthonormal basis functions, e.g., the Kautz model. Such
extensions are necessary and will be examined in our future works
because it is known that the convergence rate of the Laguerre model
is slow when the system has poles close to the unit circle. Another
interesting topic is regarding how to design a suitable kernel for
the coefficients of the orthonormal basis functions.

\section{Acknowledgement}

The authors would like to thank Prof. H{\aa}kan Hjalmarsson, Prof.
Bo Wahlberg and Asso. Prof. Cristian Rojas for their comments during
the LEARN annual meeting in 2015.







\bibliographystyle{IEEEtran}
\bibliography{../ref}

\end{document}